\documentclass[sigconf]{acmart}
\usepackage{multirow}
\usepackage{listings}

\AtBeginDocument{%
  }

\copyrightyear{2025} 
\acmYear{2025} 
\setcopyright{rightsretained}
\acmConference[ITiCSE 2025]{Proceedings of the 30th ACM Conference on Innovation and Technology in Computer Science Education V. 2}{June 27-July 2, 2025}{Nijmegen, Netherlands}
\acmBooktitle{Proceedings of the 30th ACM Conference on Innovation and Technology in Computer Science Education V. 2 (ITiCSE 2025), June 27-July 2, 2025, Nijmegen, Netherlands}\acmDOI{10.1145/3724389.3731266}
\acmISBN{979-8-4007-1569-3/2025/06}

\begin{document}

\title{AutoMCQ - Automatically Generate Code Comprehension Questions using GenAI}

\author{Martin Goodfellow}
\email{martin.h.goodfellow@strath.ac.uk}
\orcid{0000-0003-2151-8442}
\affiliation{
  \institution{University of Strathclyde}
  \city{Glasgow}
  \country{Scotland}
}

\author{Robbie Booth}
\email{robbie.booth.2021@uni.strath.ac.uk}
\orcid{0009-0002-6178-9300}
\affiliation{
  \institution{University of Strathclyde}
  \city{Glasgow}
  \country{Scotland}
  }

\author{Andrew Fagan}
\email{andrew.fagan@strath.ac.uk}
\orcid{0000-0001-9714-2096}
\affiliation{
  \institution{University of Strathclyde}
  \city{Glasgow}
  \country{Scotland}
  }

\author{Alasdair Lambert}
\email{alasdair.lambert@strath.ac.uk}
\orcid{0000-0002-9762-2193}
\affiliation{
  \institution{University of Strathclyde}
  \city{Glasgow}
  \country{Scotland}
  }

\renewcommand{\shortauthors}{Martin Goodfellow, Robbie Booth, Andrew Fagan, and Alasdair Lambert} 


\begin{abstract}
Students often do not fully understand the code they have written. This sometimes does not become evident until later in their education, which can mean it is harder to fix their incorrect knowledge or misunderstandings. In addition, being able to fully understand code is increasingly important in a world where students have access to generative artificial intelligence (GenAI) tools, such as GitHub Copilot.

One effective solution is to utilise code comprehension questions, where a marker asks questions about a submission to gauge understanding, this can also have the side effect of helping to detect plagiarism. However, this approach is time consuming and can be difficult and/or expensive to scale.

This paper introduces AutoMCQ, which uses 
GenAI for the automatic generation of multiple-choice code comprehension questions. This is integrated with the CodeRunner automated assessment platform.
\end{abstract}





\maketitle

\section{Introduction}

Students in introductory programming classes are generally able to write working code. However, they may not have a full understanding or comprehension of how the code works. 
In the experience of Lehtinen et al., around a third of their class of 125 students had difficulties explaining their code~\cite{Lehtinen21}. At this introductory stage, this can have serious long term consequences to the student's ability to become a skilled programmer, as these shaky foundations become increasingly strained by problems of greater scale and complexity. Code comprehension questions are an effective solution to this problem~\cite{schulte2010programComprehension}, forcing students to critically examine their own code, as well as challenging any incorrect assumptions.
Unfortunately, the bespoke nature - manually creating questions based on a student's own code, makes for a process which is difficult to scale to larger classes.




This work presents AutoMCQ, a tool for the automatic generation of bespoke multiple-choice code comprehension questions by utilising a combination of automated unit testing and AI generated follow-up questions. 
This is not necessarily intended as a method of assigning or gating class credit, but rather as a tool for identifying cases where there might be a benefit to early intervention.



\section{AutoMCQ} 
\begin{figure}
    \centering
    \includegraphics[width=0.25\textwidth]{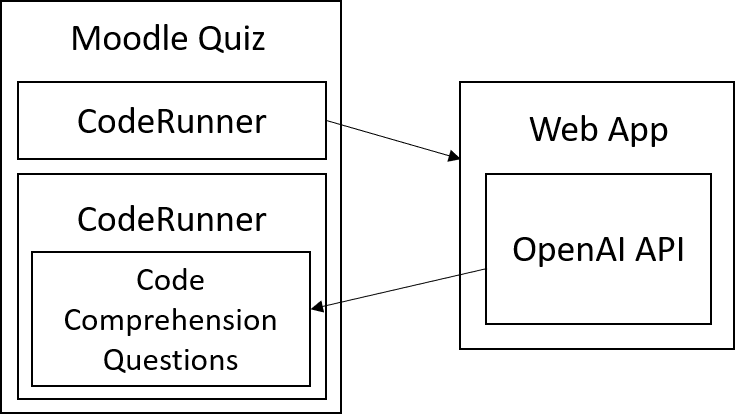}
    \caption{System Architecture}
    \label{fig:architecture}
\end{figure}

We have developed a web application, AutoMCQ, which uses GPT-4o mini via the OpenAI API to generate personalised multiple-choice code comprehension questions, based on a student's submitted code. The code for this tool is available on GitHub~\footnote{https://github.com/RobbieBooth/AI-Question-Generator}. We integrated calls to this application within CodeRunner~\cite{lobb2016} quiz questions hosted on our Virtual Learning Environment (VLE), which is built on top of Moodle~\footnote{https://moodle.org/}. CodeRunner~\footnote{https://coderunner.org.nz/} is a Moodle plugin which allows users to submit code to be run against predefined test cases and can support multiple programming languages. The high level architecture of our approach can be seen in Figure~\ref{fig:architecture}.

The students are presented with a traditional CodeRunner question (see ~\cite{Goodfellow24} for more detail). When they then progress to the code comprehension questions their code plus other parameters are passed to our web application. The prompt sent to the OpenAI API consists of the system prompt "You are an educational assistant specializing in computer science. Your task is to analyse students' code for the beginner programmer class and generate thoughtful multiple-choice questions that can help them understand and improve their coding skills. You should try and make good distractor options to really test students understanding." 
plus the parameters detailing the number of questions, CodeRunner question text, topics to ask about, programming language, any code the student was provided with, plus the student's submitted code to stop questions being asked on the skeleton code. This then returns multiple-choice code comprehension questions and displays them within a CodeRunner question. The students can then answer these questions and have them automatically marked. Unfortunately, we can't rely on GenAI generating correct or sensible questions every time. Therefore, to handle this there is a note above the generated questions: "These questions were generated by AI. Therefore, questions generated may be incorrect. If you think they are incorrect please select `This question doesn't seem right'. Also, select this option if the question doesn't relate to programming." This option can be used to trigger a manual check by the class lecturer. Until GenAI models improve our code comprehension questions will only be used for formative assessment. 




\subsection{Example}\label{sec:ex}



The following question is used in our introductory Java programming class when testing student knowledge of inheritance and overriding:

\textbf{Q:} The council tax for a flat is calculated similarly to that for a building (see Figure~\ref{fig:code}) except that there is a deduction of £75. Develop a new class called \textbf{Flat.java} which inherits from \textbf{Building.java} and correctly implements the modified \textbf{getTax()} method.

\lstset{breaklines=true}

\begin{figure}
    \centering
    \begin{lstlisting}[language=Java, basicstyle=\footnotesize]
    public class Building
    {
        private int windows;
        private double charge;

        public Building(int windows, double charge) {
            this.windows = windows;
            this.charge = charge;
        }

        public double getTax() {
            return this.windows * this.charge;
        }
    }
\end{lstlisting}
    \caption{Building Class}
    \label{fig:code}
\end{figure}

The CodeRunner tests for this question involve testing \textbf{getTax()} with different arguments and printing the answer. The result is then compared with the expected result. For example:
\begin{lstlisting}[language=Java, showstringspaces=false, basicstyle=\small]
Flat f0 = new Flat(7, 18.5);
System.out.println("Council tax for flat is: " + f0.getTax());
\end{lstlisting}

If the submitted code is correct it should output \begin{lstlisting}[basicstyle=\small]
Council tax for flat is: 54.5
\end{lstlisting}

\begin{figure}
    \centering
    \includegraphics[width=0.39\textwidth]{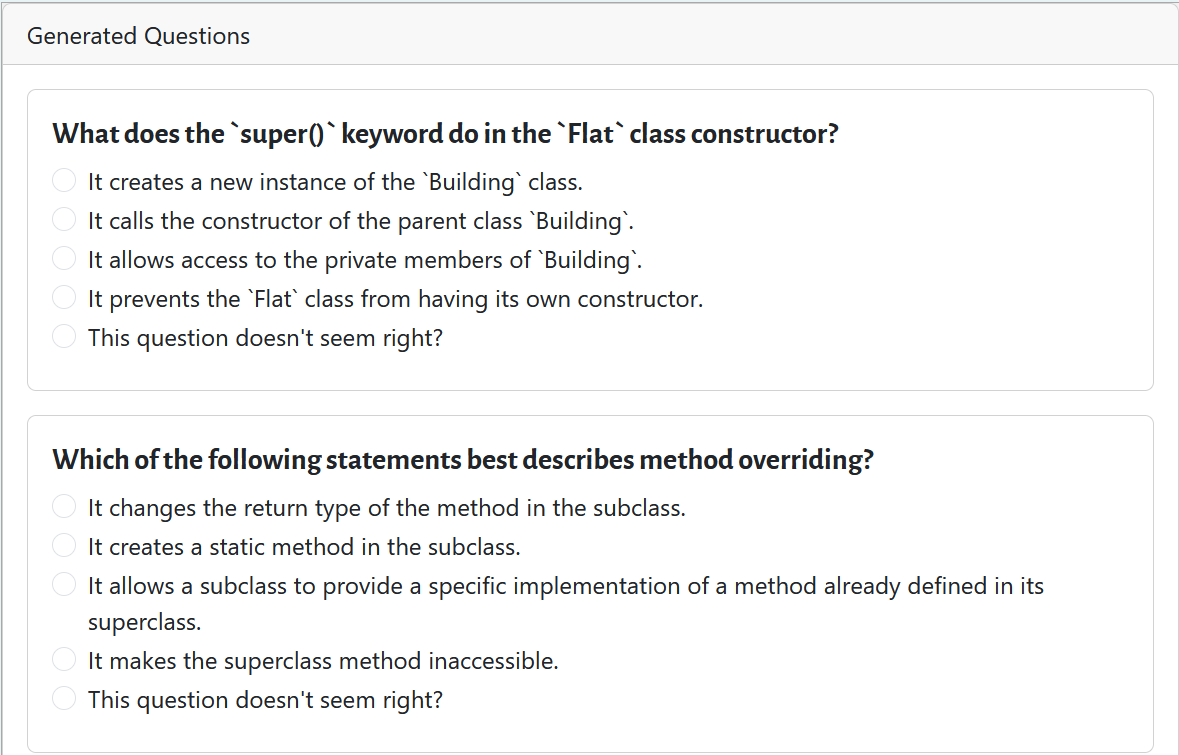}
    \caption{Generated Questions}
    \label{fig:genQs}
\end{figure}

The submitted code is sent to the web application along with number of questions 2, the question, the topics "inheritance and overriding", language java, and \textbf{Building.java}. Examples of generated questions for a correct solution can be seen in Figure~\ref{fig:genQs}.

\section{Conclusion} 
We presented our tool, AutoMCQ. This tool has received a positive response from its use as a study aid in our introductory Java programming class. In a survey, the majority of students claimed it helped them better comprehend their code. However, a larger study would be needed to formally prove this. There was also a minor increase in the average marks for the class tests but we believe this tool will have a larger impact on our advanced programming classes. However, further studies are required to confirm this.

\begin{acks}
This work was funded by Research Interns @ Strathclyde (RI@S).
\end{acks}

\bibliographystyle{ACM-Reference-Format}
\bibliography{sample-base}

\end{document}